\documentclass[twocolumn,aps,pra,superscriptaddress,showpacs,tightenlines]{revtex4}
\usepackage{amsmath}
\usepackage{amsfonts}
\usepackage{graphicx}
\usepackage{epsfig}
\usepackage{color}
\usepackage[colorlinks,citecolor=blue]{hyperref}

\begin{document}

\title{Single-Photon-Triggered Quantum Phase Transition}
\author{Xin-You L\"{u}}
\email{xinyoulu@hust.edu.cn}
\affiliation{School of physics, Huazhong University of Science and Technology, Wuhan 430074, China}

\author{Li-Li Zheng}
\affiliation{School of physics, Huazhong University of Science and Technology, Wuhan 430074, China}

\author{Gui-Lei Zhu}
\affiliation{School of physics, Huazhong University of Science and Technology, Wuhan 430074, China}

\author{Ying Wu}
\email{yingwu2@126.com}
\affiliation{School of physics, Huazhong University of Science and Technology, Wuhan 430074, China}

\begin{abstract}
We propose a hybrid quantum model combining cavity QED and optomechanics, which allows the occurrence of equilibrium superradiant quantum phase transition (QPT) triggered by a single photon. This single-photon-triggered QPT exists both in the cases of ignoring and including the so-called $A^2$ term, i.e., it is immune to the no-go theorem. It originally comes from the photon-dependent quantum criticality featured by the proposed hybrid quantum model. Moreover, a reversed superradiant QPT is induced by the competition between the introduced $A^2$ term and the optomechanical interaction. This work offers an approach to manipulate QPT with a single photon, which should inspire the exploration of single-photon quantum-criticality physics and the engineering of new single-photon quantum devices.
\end{abstract}
\maketitle

\section{Introduction}
Quantum phase transition (QPT), driven by quantum fluctuations~\cite{Sachdev2011}, is fundamentally interesting and
has potential applications in modern quantum technology~\cite{Osterloh2002,Emary2003,Lambert2004,Wang2009,Jin2014}. Dicke model (DM)~\cite{Dicke1954} predicts a superradiant QPT in the thermal equilibrium, i.e., the phase transition from a normal phase to a superradiant phase at zero temperature as increasing the spin-field interaction~\cite{Hepp1973,Wang1973,Brandes2005}, bridging the statistical physics and electrodynamics. Recent advances of quantum technology has led to the growing interest in the exploration of superradiant QPT~\cite{Li2006,Baumann2011,Baksic2014,Zou2014,Bamba2016,Kirton2017}. However the existence of this equilibrium QPT in the cavity and circuit QED systems is still under debate due to the no-go theorem induced by the so-called $A^2$ term~\cite{Rzazewski1975,Knight1978,Keeling2007,Nataf2010,Viehmann2011,Liberato2014,Vukics2014,Leib2014,Jaako2016}. Until now, the superradiant QPT has not been realized experimentally in the thermal equilibrium, while the nonequilibrium superradiant QPT~\cite{Dimer2007,Nagy2010} has been observed in the driven cold-atom system~\cite{Baumann2010,Baden2014,Klinder2015}.

Cavity optomechanics, exploring the nonlinear photon-phonon interaction, provides an alternative platform of manipulating the bosonic field at a quantum level~\cite{reviews,Lu2015,Wu2015}. In particular, the quadratic optomechanical coupling offers a photon-dependent-modulation on the phonon potential, even through it is typically very weak~\cite{Thompson2008,Bhattacharya2008,Rai2008,Sankey2010,Nunnenkamp2010,Purdy2010,Biancofiore2011,Buchmann2012,Xuereb2013,Shi2013}. Recently, it is shown that this coupling could be increased by a measurement-based method~\cite{Vanner2011}, the near-field effects~\cite{Li2012}, using a fiber cavity~\cite{Jacobs2012}, or the good tunability of superconducting circuit~\cite{Kim2015}. This expands the application prospects of quadratic optomechanics in the quantum realm~\cite{liao2013}. A natural question is whether the quadratic optomechanical coupling could influence the superradiant QPT significantly. The linear optomechanical interaction has been introduced into the cavity QED to enhance the indirect atom-phonon coupling~\cite{Hammerer2009} or obtain the rich nonlinear dynamics phenomena~\cite{Ramos2013,Restrepo2014}. However, the crossover between the quadratic optomechanics and the QPT theory remains largely unexplored, which may substantially advance the fields of cavity optomechanics and statistical physics.

Here we propose a hybrid quantum model by introducing the quadratic optomechanical coupling into a normal DM. During a parameter range $\tau$, it predicts the occurrence of equilibrium superradiant QPT triggered by a single photon of an ancillary mode. In principle, this single-photon-triggered QPT is immune to the no-go theorem, and a reversed superradiant QPT (i.e., the phase transition to superradiant phase happens as decreasing the spin-field coupling) is induced by the competition between the $A^2$ term and the quadratic optomechanical coupling. Physically, the proposed hybrid quantum model features a photon-dependent quantum criticality, which corresponds to a single-photon-induced reduction (or appearance) of quantum critical point in the case of ignoring (or including) the $A^2$ term. This ultimately leads to the single-photon-triggered QPT (together with a spontaneous $\mathbb{Z}_2$ symmetry breaking) during a parameter range $\tau$ covering the induced quantum critical point (see the following Figs.\,\ref{Fig2} and \ref{Fig4}). As far as we know, this unconventional superradiant QPT is identified for the first time, which may open the research of the single-photon-triggered quantum criticality.  

This work also has wide applications in the modern single-photon quantum technologies. On the one hand, it offers a new method to detect the single photon by measurement the excitation number of a bosonic mode. This method can effective distinguish the single photon Fock state from a coherent state with single photon amplitude. Because the latter cannot trigger the superradiant QPT deterministically. On the other hand, our work provides the theoretical basis for designing new single-photon quantum device, such as the high-precision single-photon switching.  

\section{Model}
\begin{figure}
\includegraphics[clip,width=8cm]{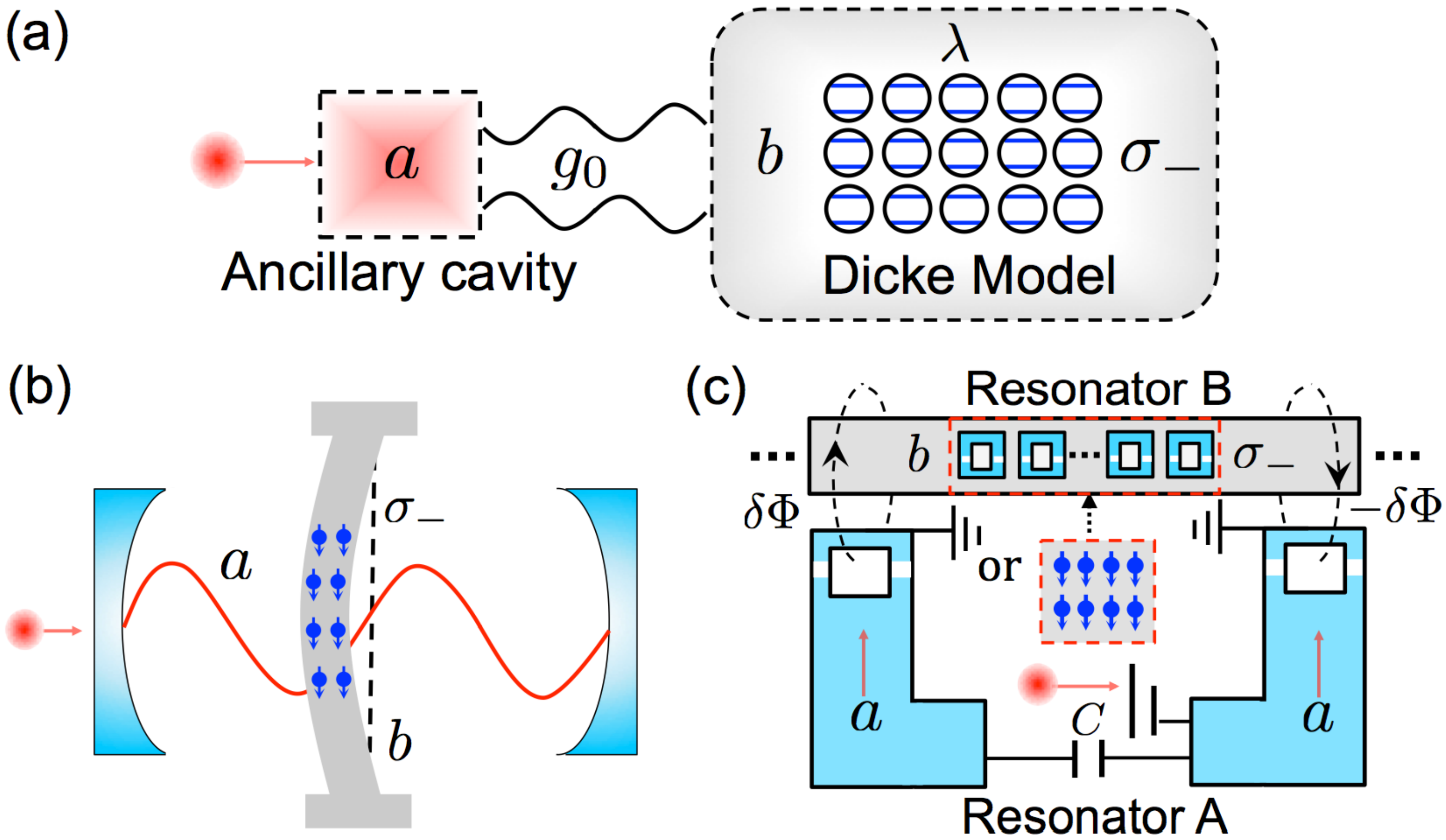}
\caption{(color online). (a) A schematic illustration of a general hybrid model for implementing single-photon-triggered QPT. It consists of a DM ($N$ two-level systems $\sigma_-$ interacting with a field mode $b$ with strength $\lambda$) quadratically coupled to an ancillary mode $a$ with strength $g_0$. The implementation of this hybrid model (b) in a quadratically coupled optomechanical system with a Òmembrane-in-the-middleÓ configuration, and (c) in a superconducting circuit with the ability of simulating a quadratic optomechanical coupling (see Ref.~\cite{Kim2015}) and coupling to the large number of superconducting qubits~\cite{Kakuyanagi2016} or spin ensemble~\cite{Kubo2011}.}
\label{Fig1}
\end{figure}

We consider a hybrid quantum model depicted in Fig.$\,$\ref{Fig1}(a) (i.e., a DM coupled to an ancillary cavity mode via a quadratic optomechanical coupling) with total Hamiltonian ($\hbar=1$)
\begin{align}
H=H_{\rm an}+H_{\rm dm}-g_{0}a^{\dagger}a(b^{\dagger}+b)^2,\label{H_or}
\end{align}
where $a$ ($a^{\dagger}$) and $b$ ($b^{\dagger}$) are the annihilation (creation) operators of the ancillary mode and the field mode of the DM, respectively. Hamiltonian $H_{\rm dm}$ is given by
\begin{align}
\!\!H_{\rm dm}=\Omega J_z+\omega b^{\dagger}b+\frac{\lambda}{\sqrt{N}}(b^{\dagger}+b)J_x+\frac{\alpha\lambda^2}{\Omega}(b^{\dagger}+b)^2,\label{H_di}\!\!
\end{align}
with a collective coupling strength $\lambda$ and the angular momentum operators $J_{z}=(1/2)\sum^N_{i=1}\sigma_z$, $J_{\pm}=\sum^N_{i=1}\sigma_\pm$, and $J_x=J_-+J_+$. It describes $N$ two-level systems $\sigma_-$ (with frequency $\Omega$) interacting with a field mode $b$ (with frequency $\omega$) and the $A^2$ term has been included in the last term. Normally $\alpha\geq1$ (decided by the Thomas-Reiche-Kuhn sum rule~\cite{Nataf2010}) corresponds to the case of implementing the DM in cavity QED system, and $H_{\rm dm}$ is reduced to the Hamiltonian of a standard DM when $\alpha=0$. The ancillary cavity, with Hamiltonian $H_{\rm an}=\omega_{c}a^{\dagger}a$, quadratically couples to $b$ with coupling strength $g_0$~\cite{Bhattacharya2008}. According to Ref.\,\cite{Bhattacharya2008}, here we consider the ancillary cavity contains an odd number of half wavelengths of mode $a$ in the full cavity, which leads to the minus sign in the quadratic coupling, i.e., the last term of Eq.\,(\ref{H_or}). Hamiltonian (\ref{H_or}) has $\mathbb{Z}_2$ symmetry associating with a well-defined parity operator $\Pi=e^{i\pi\mathcal{N}}$, where $\mathcal{N}=b^{\dagger}b+J_z+N/2$ is the total excitation number of system (excluding the ancillary mode $a$).

\begin{figure}
\includegraphics[width=8cm]{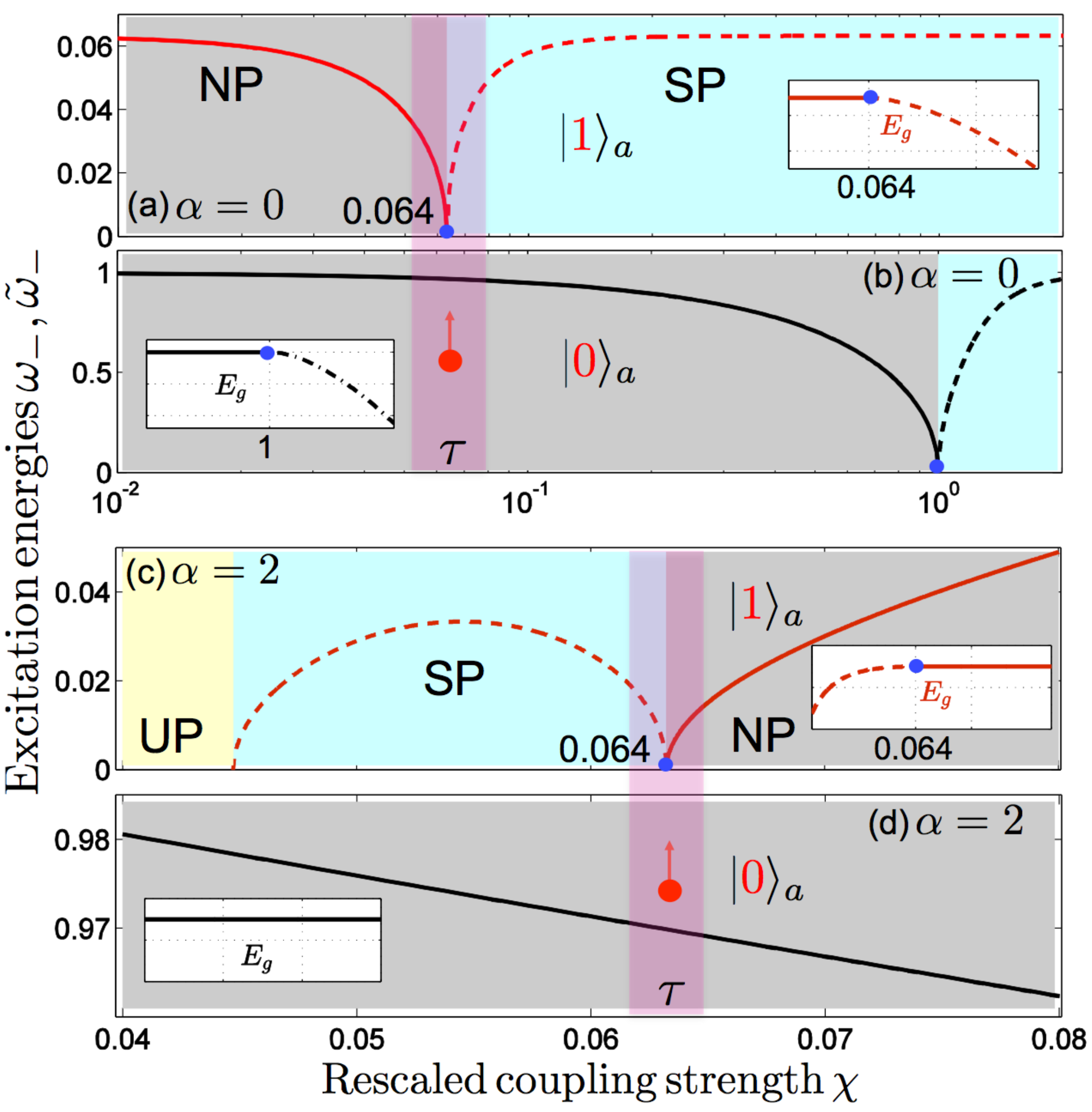}
\caption{(color online). The excitation energies $\omega_-$, $\tilde{\omega}_{-}$ and the rescaled ground-state energies $E_g/N$, $\tilde{E}_g/N$ (the inserts) versus $\chi$ when (a,b) $\alpha=0$ and (c,d) $\alpha=2$. In the figure, the normal phase, superradiant phase and unstable phase are denoted as NP, SP and UP, respectively. The blue dots indicate the quantum critical points, and the pink shading areas indicate the tunable parameter range $\tau$ used to demonstrate the single-photon-triggered QPT. The system parameters are chosen as $\Omega=\omega=1$, and (a,b) $g_0/\omega=0.249$, (c,d) $g_0/\omega=0.251$.}
\label{Fig2}
\end{figure}

The proposed hybrid quantum model could be realized in a quadratically coupled optomechanical system [see Fig.\,\ref{Fig1}(b)]. The DM can be implemented by coupling a mechanical resonator to an ensemble of two-level systems, e.g., the nitrogen-vacancy centers in diamond~\cite{Roukes2005,Arcizet2011,Bennett2013,Teissier2014}. As shown in Ref.\,\cite{Bennett2013}, to resonant with the microwave transition of the spin triplet state of NV center, the mechanical frequency requires reaching to the order of GHz. Specifically, the flex of diamond membrane strains the diamond lattice, which in turn couples directly to the spin triplet states in the nitrogen-vacancy electronic ground state. Note that this case corresponds to our following results of $\alpha=0$, since the $A^2$ term appears for electromagnetic fields normally~\cite{Rzazewski1975,Vukics2014}. Moreover, the required quadratic coupling can be realized by placing the semitransparent membrane at the middle of cavity, where $\omega'_c(0)=0$~\cite{Thompson2008,Bhattacharya2008}.

Another candidate for implementing our model is the superconducting circuit depicted in Fig.\,\ref{Fig1}(c). An ensemble of two-level systems (e.g., superconducting qubits~\cite{Kakuyanagi2016} and spin ensemble~\cite{Kubo2011}) coupled to resonator B forms the DM. Moreover, the coupling capacitor $C$ and the superconducting quantum interference devices (SQUIDs) forming resonator A offer an effective fixed semitransparent membrane and movable cavity ends, respectively. A relative displacement of the fixed membrane with respect to the center of resonator A is generated by synchronizing the motion of the moveable cavity ends, which is obtain by applying opposite flux variations $\pm\delta\Phi$ through the SQUIDs. Here $\pm\delta\Phi$ is proportional to the position quadrature $x$ of resonator B. This ultimately leads to a $x$-dependent frequency $\omega_c(x)$ of resonator A. Now the position quadrature $x$ effectively oscillates in the middle of resonator A, where $\omega'_c(0)=0$. The quadratic optomechanical coupling between resonators A and B is realized~\cite{Kim2015}.
In addition, our proposal also might be implemented in cavity opto-mechanic setup with a BEC~\cite{Brennecke2008}.

\begin{figure}[t]
\centerline{\includegraphics[clip,width=8cm]{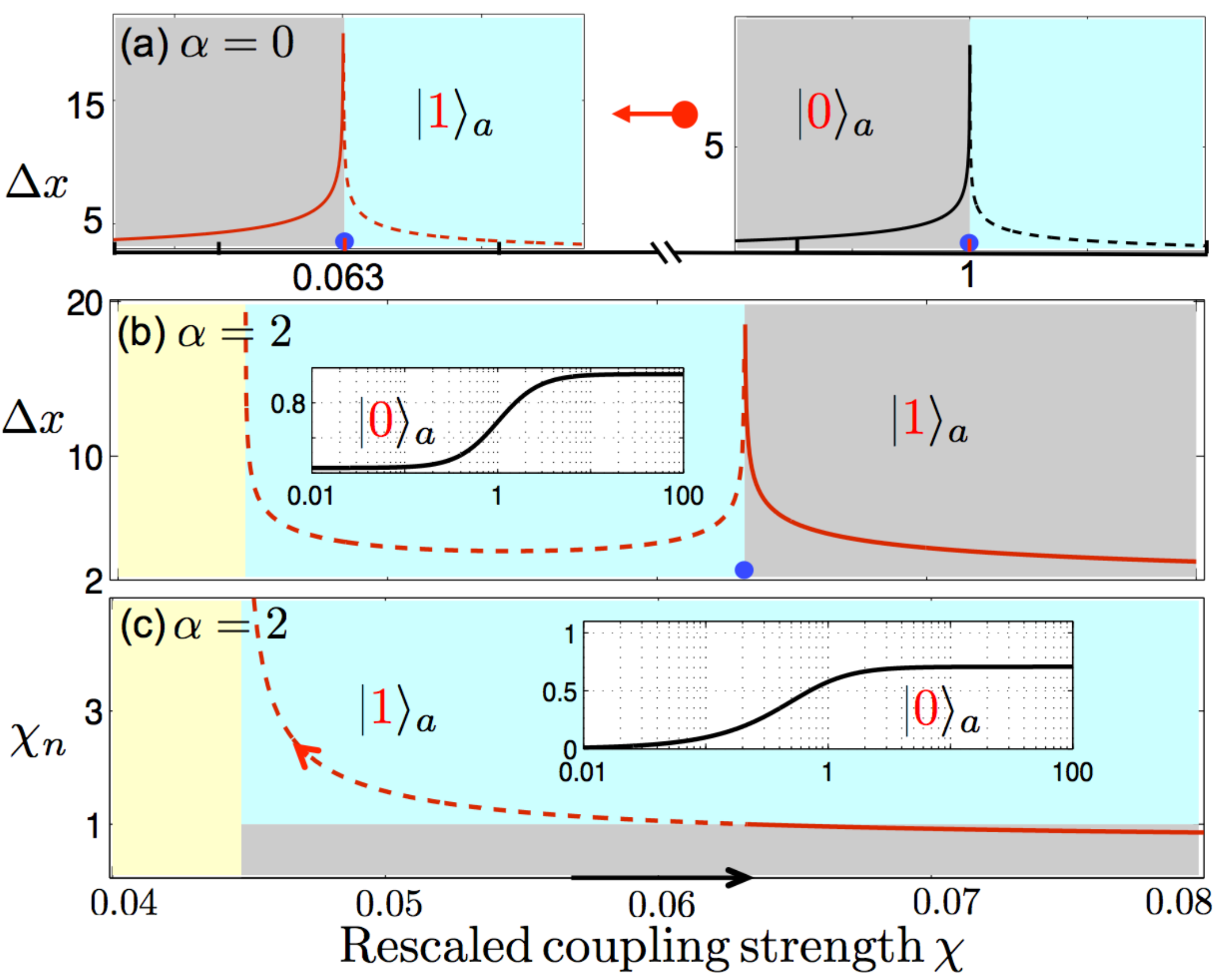}}
\caption{(color online). The ground-state position-variance $\Delta x$ versus coupling strength $\chi$ when (a) $\alpha=0$ and (b) $\alpha=2$. Here $\Delta x=\sqrt{\langle x^2\rangle_g-\langle x\rangle_g^2}$ with $x=(1/\sqrt{2})(b^{\dagger}+b)$, and the blue dots denote the corresponding singular points. (c) The photon-dependent characteristic parameter $\chi_n$ versus coupling strength $\chi$. The main and insert parts of (b,c) denote when the ancillary mode is in state $|1\rangle_a$ and $|0\rangle_a$, respectively.}
\label{Fig3}
\end{figure}
\begin{figure}[t]
\centerline{\includegraphics[clip,width=8cm]{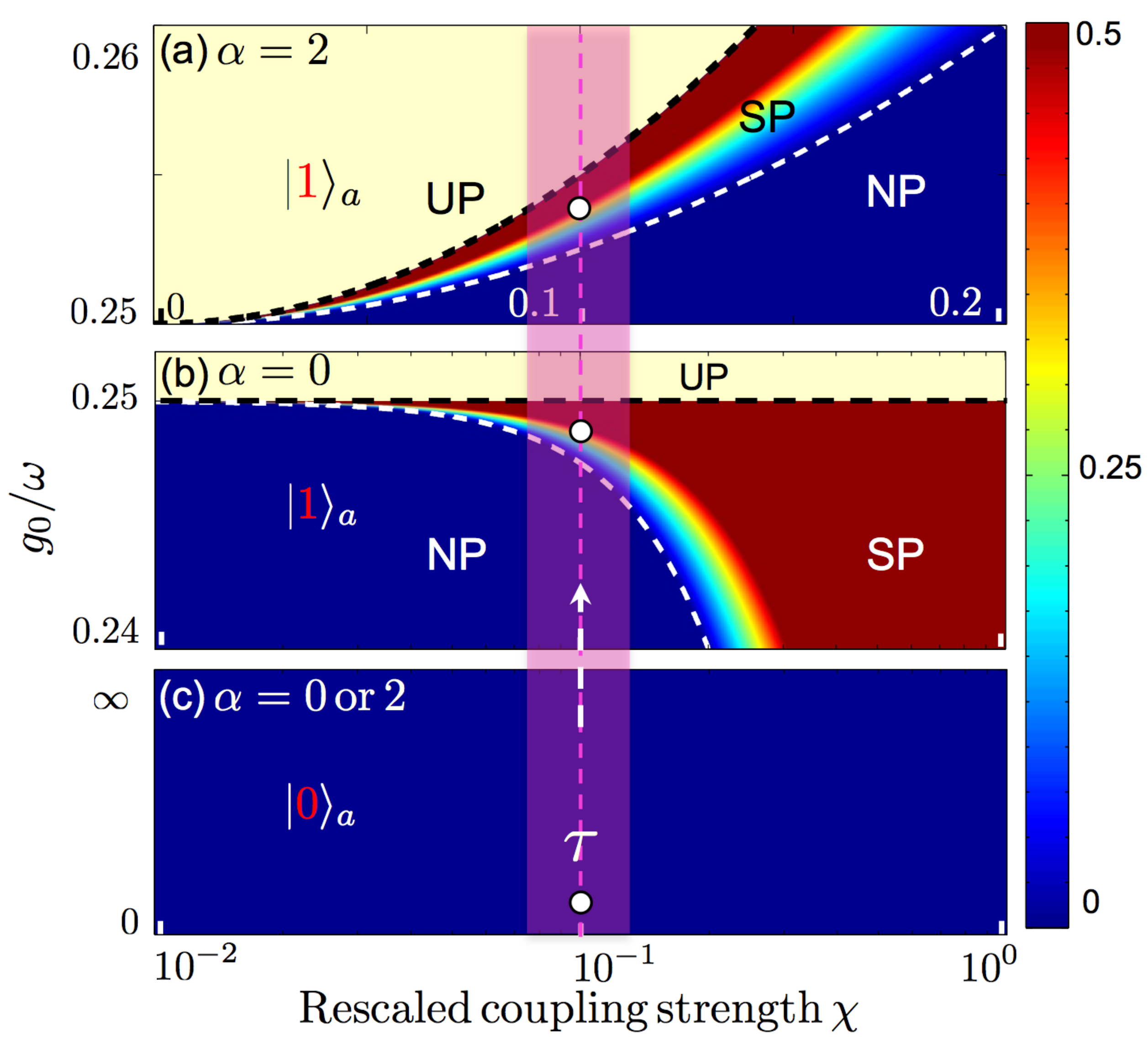}}
\caption{(color online). Photon-dependent ground-state phase-diagram. The order parameter $\psi_q$ versus $\chi$ and $g_0$ when (a,b) $n=1$ and (c) $n=0$. The NP and SP correspond to the region of $\psi_q=0$ and $\psi_q>0$, respectively. The white (black) dashed contour indicates the position where $\psi_q$ becomes non-zero (an imaginary number), locating the phase transition from NP to SP (SP to UP). The shading area $\tau$ denote any of ranges showing the single-photon-triggered QPT.}
\label{Fig4}
\end{figure}

\section{Photon-dependent quantum criticality}

The optomechanical interaction in Hamiltonian (\ref{H_or}) provides a photon-dependent modification on the potential of field $b$. It is different from the original $A^2$ term, whose strength is altered when one checks the occurrence of QPT by changing $\lambda$. Considering the ancillary mode $a$ is prepared into the Fock state $|n\rangle_a$ ($n=0,1,...$), the number operator $a^{\dagger}a$ can be replaced by an algebraic number $n$. Note that, this replacement is not  suited to the coherent state with single photon amplitude due to the coherent state is not the eigenstate of $a^{\dagger}a$. Then, applying a squeezing transformation $b=\cosh(r_n)b_n+\sinh(r_n)b_n^{\dagger}$ with $r_n=(-1/4)\ln[1+\alpha\chi^2-4ng_0/\omega]$ and a rescaled coupling strength $\chi=2\lambda/\sqrt{\Omega\omega}$, the Hamiltonian (\ref{H_or}) becomes
\begin{equation}
H_{n}=\Omega J_z+\omega_n b_n^{\dagger}b_n+\frac{\lambda_n}{\sqrt{N}}(b_n^{\dagger}+b_n)J_x+C_n,
\label{H_n}
\end{equation}
where $\omega_n=\exp(-2r_n)\omega$, $\lambda_n=\exp(r_n)\lambda$ and $C_n=n\omega_c+[\exp(-2r_n)-1](\omega/2)$. It clearly shows that the present hybrid quantum model is essentially equivalent to a photon-dependent DM. It is stable only when $\omega_n\geq0$, which leads to the stable conditions of system $g_0\leq\omega/(4n)$ and $\chi\geq\sqrt{(1/\alpha)(4ng_0/\omega-1)}$ for the case of ignoring and including the $A^2$ term (i.e., $\alpha=0$ and $\alpha\geq1$), respectively. The system enters into a unstable phase when $\omega_n<0$, i.e., the yellow areas of Figs.\,\ref{Fig2}-\ref{Fig4}.

In the thermodynamic limit $N\rightarrow\infty$, the system can be diagonalized analytically (see Appendix \ref{app}). A photon-dependent quantum critical point, $\chi_n=2\lambda_n/\sqrt{\Omega\omega_n}=1$ is obtained, corresponding to $\chi=\exp(-2r_n)=\sqrt{1+\alpha\chi^2-4ng_0/\omega}$ in term of the original system parameters. When
$\chi<\exp(-2r_n)$ the system is in the normal phase, featured by a lowest excitation energy $\omega_{-}$ and an rescaled ground-state-energy $E_g/N=-\Omega/2$. The corresponding ground state of system is a two-mode-oscillator ground-state $|G\rangle_{\rm np}=|00\rangle_e$ with $e^{\dagger}_{i}e_{i}|00\rangle_e=0|00\rangle_e$ $(i=1,2)$, and it has a conserved $\mathbb{Z}_2$ symmetry, testified by the zero ground-state coherence of field $\langle b\rangle_g=0$. The excitation energy $\omega_{-}$ vanishes when $\chi=\exp(-2r_n)$, locating the superradiant QPT. When $\chi>\exp(-2r_n)$, the system enters into the superradiant phase and has a lowest excitation energy $\tilde{\omega}_{-}$ together with an rescaled ground-state-energy $\tilde{E}_g/N=-(\Omega/4)(\chi_n^2+\chi_n^{-2})$. Now the ground state $|G\rangle^{\pm}_{\rm sp}$ becomes twofold degenerate and it corresponds to a spontaneous $\mathbb{Z}_2$ symmetry breaking, as is evident from the non-zero ground-state coherence of field $\langle b\rangle^{\pm}_{g}=\pm \exp(r_n)\beta$ (here the sign $\pm$ is used to distinguish two degenerate ground state). Moreover, the rescaled ground-state occupation of field $b$ can be defined as the order parameter charactering superradiant QPT, i.e., $\psi_{\rm q}=[\exp(-4r_n)\omega/(N\Omega)]\langle b^{\dagger}b\rangle_g$. Then $\psi_{q}=0$ when $\chi<\exp(-2r_n)$, and $\psi_{\rm q}=(1/4)(\chi_n^2-\chi_n^{-2})$ becomes non-zero when $\chi>\exp(-2r_n)$.

\section{Single-photon-triggered superradiant QPT}
Interestingly, the photon-dependent quantum criticality featured in our model leads to a single-photon-triggered QPT, when we focus on the case of $n=0,1$. Specifically, when the ancillary mode $a$ is in the vacuum state $|0\rangle_a$, Hamiltonian (\ref{H_n}) is reduced to a standard Dicke Hamiltonian. The superradiant QPT occurs at $\chi=1$ when $\alpha=0$ and it is prevented when $\alpha\geq1$ due to the no-go theorem. When $a$ is in the single-photon state $|1\rangle_a$, the superradiant QPT occurs at $\chi=\exp(-2r_1)=\sqrt{1+\alpha\chi^2-4g_0/\omega}$, which could be much smaller than 1 for both the case of $\alpha=0$ and $\alpha\geq1$, by properly choosing system parameters. Let's consider a parameter range $\chi\in\tau$ to check the occurrence of superradiant QPT ($\tau$ covering the single-photon-induced quantum critical point, i.e., $\chi=e^{-2r_1}$). As shown in Fig.\,\ref{Fig2}, the superradiant QPT during $\tau$ is triggered by exciting a single photon in mode $a$ (i.e., $|0\rangle_a\rightarrow|1\rangle_a$), named as ``single-photon-triggered QPT''. It corresponds to a single-photon-triggered $\mathbb{Z}_2$ symmetry breaking, demonstrated by the ground-state coherence of field $\langle b\rangle_g$.
\begin{figure}[t]
\centerline{\includegraphics[clip,width=8cm]{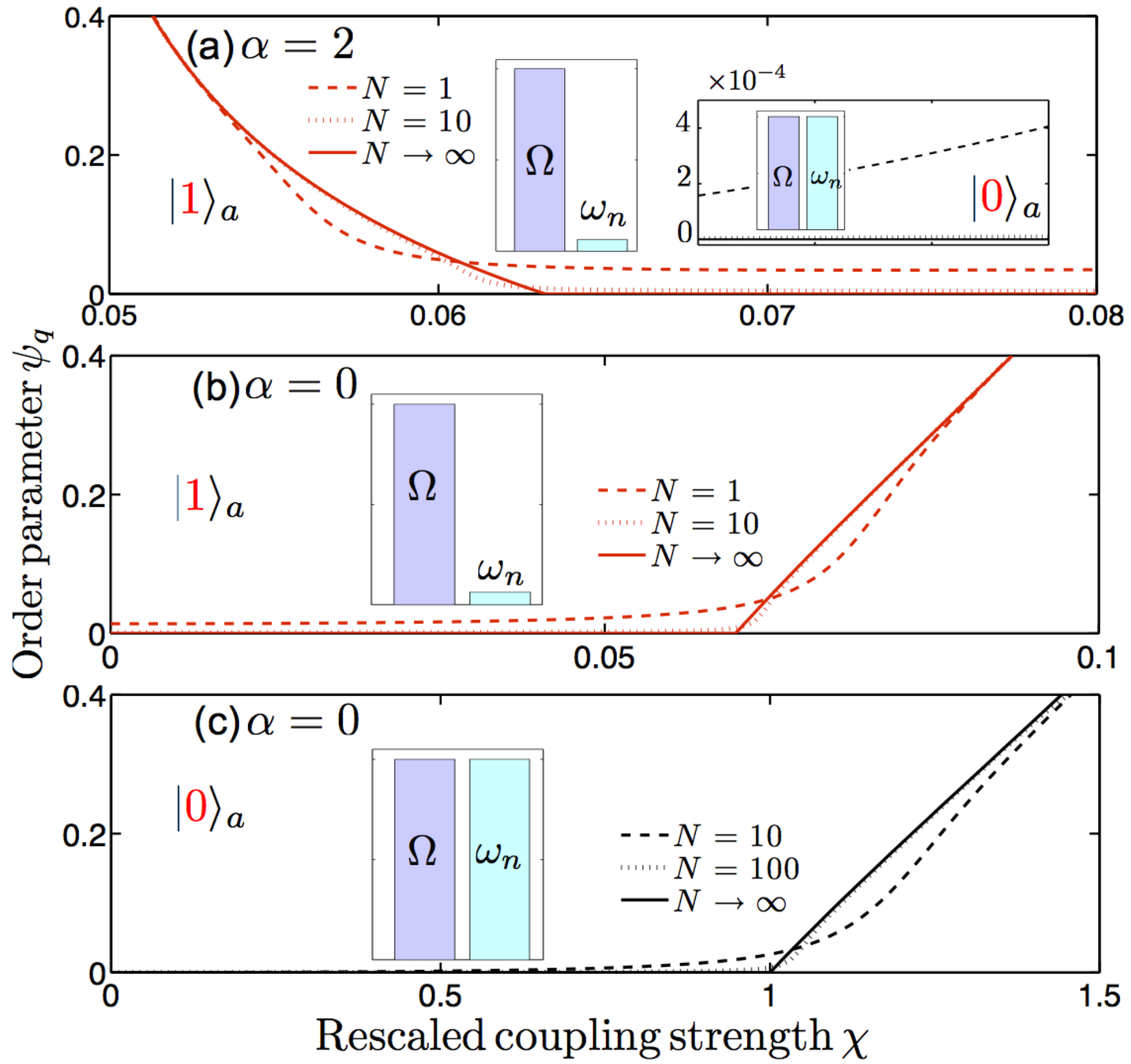}}
\caption{(color online). (a) The order parameter $\psi_q$ versus $\chi$ for different $N$ when $\alpha=2$. The main and inserted plots correspond to the case of $n=1$ and $0$, respectively. (b,c) The order parameter $\psi_q$ versus $\chi$ when $\alpha=0$ and the ancillary mode $a$ is in (b) $|1\rangle_a$ and (c) $|0\rangle_a$. All the inserted bar graphs clearly present the photon-dependent ratio $\Omega/\omega_n$.}
\label{Fig5}
\end{figure}

Including the $A^2$ term, this superradiant QPT can still occur, since the parameter condition of QPT, $\chi>\sqrt{1+\alpha\chi^2-4g_0/\omega}$ can be satisfied even when $\alpha\geq1$. Moreover, the present superradiant QPT is reversed comparing with the case happened in a standard DM, i.e., the transition from the normal phase to the superradiant phase occurs as decreasing the original system parameter $\chi$, as shown in Figs.\,\ref{Fig2}(c) and \ref{Fig4}(a). Physically, Hamiltonian (\ref{H_n}) clearly show that the phase transition to superradiant phase occurs when $\chi_n$ is increased to larger than 1. However, in our model, the competition between the quadratic optomechanical coupling and the $A^2$ term induces the result that $\chi_n$ increases along with decreasing $\chi$, as shown in Fig.\,\ref{Fig3}(c). This ultimately leads to the reversed superradiant QPT in terms of $\chi$.

Note that, the proposed ``single-photon-triggered QPT'' occurs as changing system parameter $\chi$ during $\tau$ at the zero temperature and its essence is still the quantum fluctuation, which leads to the singularity of ground-state position-variance $\Delta x$ at quantum critical point. As shown in Figs.\,\ref{Fig3}(a,b), in our mode, a single photon can induce the dramatically reducing (or appearance) of singular point in $\Delta x$ when the $A^2$ term is ignored (or included). These singular points corresponds exactly to the quantum critical point locating the QPT.

In Fig.\,\ref{Fig4}, we plot the ground-state phase-diagram charactered by $\psi_q$, which presents a rich photon-dependent dynamics. Firstly, the single-photon-triggered superradiant QPT is exhibited during any parameter range $\tau$ covering the quantum critical point (i.e., the white dashed line given by $\chi=e^{-2r_1}$). This quantum critical point continuously decreases along with $g_0\rightarrow\omega/4$ from $g_0<\omega/4$ (or $g_0>\omega/4$) when the $A^2$ term is ignored (or included). In principle, $\tau$ could be in a weak coupling regime, i.e., $\chi\ll1$ or $\lambda\ll\sqrt{\Omega\omega}/2$. Secondly, the system becomes unstable as increasing $g_0$ or decreasing $\chi$, when $\psi_q$ becomes an imaginary number corresponding to $\omega_n<0$ (see the unstable phase denoted by the yellow area). Lastly, when fixing $\chi$, a macroscopic ground-state occupation of field $b$ could be induced by injecting a single photon into cavity $a$ (see the white dots and dashed arrow). While this transition does not belong to the standard QPT, it might has wide applications in the high-precision single-photon detection and the engineering of new single-photon quantum devices.

In Fig.\,\ref{Fig5}, we present the dependence of $\psi_q$ on $\chi$ in the finite $N$, which clearly approaches the case of $N\rightarrow\infty$ limit (i.e., the solid lines) with increasing $N$. Comparing with a normal DM [being equivalent to the case of $n=0$, i.e., Fig.\,\ref{Fig5}(c)], this tendency could be faster in our model triggered by a single photon, as shown in Figs.\,\ref{Fig5}(a,b). This might relax the parameter limit of approximately implementing superradiant QPT experimentally. For example, the superradiant QPT could be approximately demonstrated under the condition of $N\geq10$ in our model [see Figs.\,\ref{Fig5}(a,b)], while it requires $N\geq100$ in the normal DM without including the $A^2$ term [see Fig.\,\ref{Fig5}(c)]. Physically, in our model, a single photon can induce a dramatically increasing of the ratio $\Omega/\omega_n$ (see the bar graphs in Fig.\,\ref{Fig5}). This leads to the result that, comparing with a normal DM, our model can be closer to the classical oscillator limit (i.e., $\Omega/\omega_n\rightarrow\infty$), in which the superradiant QPT occurs exactly~\cite{Bakemeier2012,Hwang2015,Liu2017}.

\section{Discussion}
To implement our proposal in the optomechanical system, there still exists experimental challenge with current accessible technology. The key challenge is to reach $g_0\approx\omega/4$ ($g_0\approx\omega''_c(0)x_{\rm zpf}^2$ and $x_{\rm zpf}$ being the mechanical zero-point fluctuation). Based on current experiments~\cite{Thompson2008,Sankey2010}, for a mechanical beam with 50 pg, one can choose $\omega/2\pi\sim$100 kHz, $\omega''_c(0)\sim4.5$ MHz/nm$^2$, and $x_{\rm zpf}\sim0.5$ pm, leading to $g_0\sim$ Hz. Then to reach $g_0\approx\omega/4$, it needs to dramatically enhance $\omega''_c(0)$ and $x_{\rm zpf}$ in the future advanced experiments. Many methods have been proposed to enhance this quadratic coupling, such as employing the near-field effects~\cite{Li2012} or a fiber cavity~\cite{Jacobs2012}. 

Moreover, the flex of membrane strains the diamond lattice, which in turn couples directly to the spin ensemble via an effective strain-induced electric field. This system with strain-induced spin-phonon interaction can be described by $H_{\rm dm}$ with $\alpha=0$ (i.e., no quadratic mechanical potential)~\cite{Maze2011,Doherty2012}. By properly locating the spin ensemble near the surface of the membrane with optimized Young's modulus, $\lambda$ can reach the order of kHz~\cite{Bennett2013,Zhang2015,Xia2016} corresponding to $\chi\sim0.02$, which reaches the order of realizing QPT required here. Employing the squeezing effect into the sideband cooling technology, the low-frequency mechanical oscillators can be cooled to its ground state with thermal occupancy of 0.19 phonons~\cite{Clark2017}, which will not destroy our QPT charactered by a mechanical macroscopic excitation. However the quantum noise associating with the mechanical decay rate $\gamma_m$ (from Hz to kHz~\cite{reviews}) will influence this macroscopic excitation during the reading process. It requires that the reading time should be much smaller than $1/\gamma_m$ (from second to millisecond). Moreover, one should implement the QPT after completing the mechanical cooling to reduce the influence from the quantum noise associating with an optical decay rate. 

The superconducting circuit, depicted in Fig.\,\ref{Fig1}(c), is a promising candidate for implementing our proposal.
First, $g_0$ can be enhanced dramatically by optimizing the coupling capacitance $C$ and the tunable bias flux through the SQUIDs~\cite{Kim2015}. Secondly, the collective coupling strength $\lambda\sim100$ MHz can be realized in the superconducting circuit~\cite{Kakuyanagi2016},  leading to $\chi$ reaching the order of $10^{-2}$ required here. Lastly, regarding the description of circuit QED system with $N$ artificial atoms, there still exists debate on the $A^2$ term originally from the effective kinetic energy of system~\cite{Nataf2010,Viehmann2011}. However, as shown the above results, the proposed single-photon-triggered QPT can be realized both in the cases of ignoring and including the $A^2$ term.

\section{Conclusion}
In summary, we have proposed a single-photon-triggered superradiant QPT (associating with a single-photon-triggered $\mathbb{Z}_2$ symmetry breaking) by combining cavity QED and optomechanics. We showed that this superradiant QPT will not be limited by the no-go theorem, and the competition between the $A^2$ term and the quadratic optomechanical coupling induces a reversed superradiant QPT in our model. Moreover, this superradiant QPT also could be implemented approximately in a finite parameter regime (e.g., $N\geq10$). This work may fundamentally broaden the fields of cavity QED, optomechanics and statistical physics. It offers the prospect of exploring the single-photon-triggered quantum criticality together with its applications in the high-precision single-photon quantum technologies.

\begin{acknowledgements}
This work is supported by the National Key Research and Development Program of China grant 2016YFA0301203, the National Science Foundation of China (Grant Nos. 11374116, 11574104 and 11375067).
\end{acknowledgements}

\appendix

\section{Diagonalization procedure of Hamiltonian (\ref{H_n})\label{app}}

To diagonalize Hamiltonian (\ref{H_n}), we introduce a bosonic mode $d$ ($[d,d^{\dagger}]=1$) by using the Holstein-Primakoff representation of the angular momentum operators, i.e., $J_{+}=d^{\dagger}\sqrt{N-d^{\dagger}d}$, $J_-=\sqrt{N-d^{\dagger}d}d$, and $J_z=d^{\dagger}d-N/2$. In terms of $d$, the Hamiltonian (\ref{H_n}) becomes
\begin{align}
\!\!H_{n}=&\,\,\Omega(d^{\dagger}d-N/2)+\omega_n b_n^{\dagger}b_n\nonumber
\\
\!\!&+\lambda_n(b_n^{\dagger}\!+\!b_n)\!\!\left(d^{\dagger}\sqrt{1\!-\!\frac{d^{\dagger}d}{N}}\!+\!\sqrt{1\!-\!\frac{d^{\dagger}d}{N}}d\right)\!+\!C_n.\!
\label{H_me}
\end{align}
In the thermodynamic limit $N\rightarrow\infty$, this Hamiltonian could be reduced to a bilinear formation when $\chi<\exp(-2r_n)$ (corresponding to the  normal phase),
\begin{align}
H_{\rm np}=&\,\,\Omega d^{\dagger}d+\omega_nb_n^{\dagger}b_n\nonumber
\\
&+\lambda_n(d^{\dagger}+d)(b^{\dagger}_n+b_n)-\frac{N}{2}\Omega+C_n,
\end{align}
by ignoring terms with powers of $N$ in the denominator. Then Hamiltonian $H_{\rm np}$ can be diagonalized to $H_{\rm np}=\omega_{-} {e}_{1}^{\dagger}e_{1}+\omega_{+}{e}_{2}^{\dagger}e_{2}+E_{g}$ via a Bogoliubov transformation given by
\begin{subequations}
\begin{align}
b_n=\,&\xi^{(b)}_{-}e^{\dagger}_{1}+\xi^{(b)}_{+}e_{1}+\zeta^{(b)}_{-}e^{\dagger}_{2}+\zeta^{(b)}_{+}e_{2},
\\
d=\,&\xi^{(d)}_{-}e^{\dagger}_{1}+\xi^{(d)}_{+}e_{1}+\zeta^{(d)}_{-}e^{\dagger}_{2}+\zeta^{(d)}_{+}e_{2},
\end{align}
\end{subequations}
and their Hermitian conjugations. The coefficients are given by
\begin{subequations}
\begin{align}
\xi^{(b)}_{\pm}=&\frac{\cos\theta(\omega_n\pm\omega_{-})}{2\sqrt{\omega_n\omega_{-}}},\zeta^{(b)}_{\pm}=\frac{\sin\theta(\omega_n\pm\omega_{+})}{2\sqrt{\omega_n\omega_{+}}},
\\
\xi^{(d)}_{\pm}=&-\frac{\sin\theta(\Omega\pm\omega_{-})}{2\sqrt{\Omega\omega_{-}}}, \zeta^{(d)}_{\pm}=\frac{\cos\theta(\Omega\pm\omega_{+})}{2\sqrt{\Omega\omega_{+}}}.
\end{align}
\end{subequations}
Here the angle $\theta$ is decided by $\tan(2\theta)=4\lambda_n\sqrt{\Omega\omega_n}/(\Omega^2-\omega^2_n)$. Here the excitation energies
\begin{align}
\omega^{2}_{\pm}=\frac{1}{2}\left[\omega_n^2+\Omega^2\pm\sqrt{(\omega^2_n-\Omega^2)^2+4\chi^2\Omega^2\omega^2}\right],
\end{align}
and the rescaled ground-state-energy $E_g/N=-\Omega/2$. The corresponding ground state of system is a two-mode oscillator ground-state $|G\rangle_{\rm np}=|00\rangle_e$ defined by $e^{\dagger}_{i}e_{i}|00\rangle_e=0|00\rangle_e$ $(i=1,2)$. It has a conserved parity symmetry i.e., $\Pi|G\rangle_{\rm sp}=|G\rangle_{\rm sp}$, testified by the zero ground-state coherence of field $\langle b\rangle_g=0$. Here $\langle b\rangle_g$ can be calculated by using the relationship between modes $b$ and $e_{i}$ given by the used transformations, including the squeezing transformation $b\rightarrow b_n$ and the Bogoliubov transformation $b_n\rightarrow e_{i}$.

When $\chi>\exp(-2r_n)$ (i.e., in the superradiant phase), both the field $b_n$ and the two-level systems are macroscopically excited, which leads to the result that Hamiltonian $H_{\rm np}$ becomes invalid. In this case, we firstly apply one of two displacements into modes $b_n$ and $d$ with amplitudes $\beta=\sqrt{\frac{\Omega N}{4\omega_n}(\chi^2_n-\chi^{-2}_n)}$ and $\nu=\sqrt{\frac{N}{2}(1-\chi^{-2}_n)}$, i.e., $b_n\rightarrow\tilde{b}_n+\beta$, $d\rightarrow\tilde{d}-\nu$ or $b_n\rightarrow\tilde{b}_n-\beta$, $d\rightarrow\tilde{d}+\nu$. By the similar procedure used to derive $H_{\rm np}$, Hamiltonian (\ref{H_me}) is reduced to another bilinear Hamiltonian $H_{\rm sp}$ in terms of $\tilde{b}_n$ and $\tilde{d}$, which can be diagonalized to $H_{\rm sp}=\tilde{\omega}_{-}\tilde{e}_{1}^{\dagger}\tilde{e}_{1}+\tilde{\omega}_{+}\tilde{e}_{2}^{\dagger}\tilde{e}_{2}+\tilde{E}_{g}$ via a Bogoliubov transformation given by
\begin{subequations}
\begin{align}
\tilde{b}_n=\,\,&\tilde{\xi}^{(b)}_{-}\tilde{e}^{\dagger}_{1}+\tilde{\xi}^{(b)}_{+}\tilde{e}_{1}+\tilde{\zeta}^{(b)}_{-}\tilde{e}^{\dagger}_{2}+\tilde{\zeta}^{(b)}_{+}\tilde{e}_{2},
\\
\tilde{d}=\,\,&\tilde{\xi}^{(d)}_{-}\tilde{e}^{\dagger}_{1}+\tilde{\xi}^{(d)}_{+}\tilde{e}_{1}+\tilde{\zeta}^{(d)}_{-}\tilde{e}^{\dagger}_{2}+\tilde{\zeta}^{(d)}_{+}\tilde{e}_{2},
\end{align}
\end{subequations}
and their Hermitian conjugations. The coefficients are given by
\begin{subequations}
\begin{align}
\tilde{\xi}^{(b)}_{\pm}=&\frac{\cos\tilde{\theta}(\omega_n\pm\tilde{\omega}_{-})}{2\sqrt{\omega_n\tilde{\omega}_{-}}}, \zeta^{(b)}_{\pm}=\frac{\sin\tilde{\theta}(\omega_n\pm\tilde{\omega}_{+})}{2\sqrt{\omega_n\tilde{\omega}_{+}}},
\\
\tilde{\xi}^{(d)}_{\pm}=&-\frac{\sin\tilde{\theta}(\tilde{\Omega}\pm\tilde{\omega}_{-})}{2\sqrt{\tilde{\Omega}\tilde{\omega}_{-}}}, \tilde{\zeta}^{(d)}_{\pm}=\frac{\cos\tilde{\theta}(\tilde{\Omega}\pm\tilde{\omega}_{+})}{2\sqrt{\tilde{\Omega}\tilde{\omega}_{+}}}.
\end{align}
\end{subequations}
Here the angle $\tilde{\theta}$ is decided by $\tan(2\tilde{\theta})=2\omega_n\Omega/(\chi^4_n\Omega^2-\omega^2_n)$ and $\tilde{\Omega}=\Omega(1+\chi^2_n)/2$.
The excitation energies
\begin{align}
\tilde{\omega}^2_{\pm}=\frac{1}{2}\left[\omega^2_n+\chi_n^4\Omega^2\pm\sqrt{(\chi_n^4\Omega^2-\omega_n^2)^2+4\Omega^2\omega^2_n}\right]
\end{align}
and the rescaled ground-state-energy $\tilde{E}_g/N=-(\Omega/4)(\chi_n^2+\chi_n^{-2})$. Now the ground state of system $|G\rangle^{\pm}_{\rm sp}=|00\rangle^{\pm}_e$ with $\tilde{e}_{i}^{\dagger}\tilde{e}_{i}|00\rangle^{\pm}_e=0|00\rangle^{\pm}_e$ $(i=1,2)$ becomes twofold degenerate, i.e., the same ground state energy $\tilde{E}_{g}$ is obtained based on the above two choices of displacement applied into the modes $b_n$ and $d$. Here the sign $\pm$ corresponds to the direction of displacement applied into $b_n$, which ultimately leads to different relationship between the original field $b$ and the introduced mode $\tilde{e}_{i}$. Consequently, the $\mathbb{Z}_2$ symmetry of this ground state is spontaneously broken, as is evident from the non-zero ground-state coherence of field $\langle b\rangle^{\pm}_{g}=\pm \exp(r_n)\beta$, which could be calculated by using the relationship between modes $b$ and $\tilde{e}_{i}$ given by all used transformations, including the squeezing transformation $b\rightarrow b_n$, the displacement transformation $b_n\rightarrow \tilde{b}_{n}$, and the Bogoliubov transformation
$\tilde{b}_n\rightarrow \tilde{e}_{i}$.

\end{document}